\documentstyle[preprint,prd,aps,psfig]{revtex}\tighten
\newcommand{\beq}{\begin{equation}}
\newcommand{\eeq}{\end{equation}}
\newcommand{\beqa}{\begin{eqnarray}}
\newcommand{\eeqa}{\end{eqnarray}}
\newcommand{\k}{\kappa}

\newcommand{\siml}{\lesssim}
\begin{document}
\widetext
\draft
\preprint{\tighten\vbox{\hbox{KUNS-1718}\hbox{astro-ph/0106550}}}

\title{
Extended Quintessence and its Late-time Domination
}
\author{Takeshi Chiba}

\address{
Department of Physics, Kyoto University,
Kyoto 606-8502, Japan}

\date{April 18, 2001}

\maketitle

\bigskip

\begin{abstract}
Various astronomical observations point towards the evidence for dark energy. 
One of the most mysterious problem is the coincidence 
problem: why dark energy becomes dominant only recently. 
We present a scenario based on extended quintessence 
models to explain the late-time domination of dark energy without severe 
fine-tuning of initial conditions and model parameters. 

\end{abstract}


\section{Introduction}

The evidence for dark energy seems now compelling \cite{obs}. 
Dark energy may be 
the cosmological constant or quintessence (dynamical lambda). 
In any case, very mysterious is the fact that dark energy appears to 
become dominant only recently, although it should be negligible for the 
most of the history 
of the universe for the success of the standard big-bang nucleosynthesis 
and structure formation: we live in a very special time when the dark 
energy density is comparable to the matter density.
The coincidence (or ``why now'') problem \cite{turner} 
is the enigma  in modern cosmology.

In order to ``solve'' the problem dynamically, the dark energy density
should scale in the same way as the radiation density during 
the radiation dominated epoch; otherwise it is nothing but introducing 
a fine-tuning to account 
for the coincidence from the very beginning, 
and it is no surprise that there is some epoch when the two energy 
components coincide (however, see \cite{ewan}). 
On the other hand, however, during matter dominated epoch, dark energy 
should not track matter; otherwise dark energy cannot dominate.

Several such dynamical approaches to solving the coincidence problem have 
been attempted which utilize non-canonical kinetic terms \cite{rw,ams}
(see also \cite{coy}) or 
the explicit coupling between  quintessence and dark matter 
\cite{coupled,at}. In this paper, we shall provide yet another mechanism 
based on extended quintessence models \cite{chiba,extended}.

\section{Extended Quintessential Approach to the Coincidence Problem}

\paragraph*{Basic Idea.}

We consider the cosmological dynamics described by the 
action \cite{chiba,extended}
\beq
S=\int d^4x\sqrt{-g}\left[{1\over 2\k^2}{\cal R}-F(\phi){\cal R}-
{1\over 2}(\nabla \phi)^2-V(\phi)\right]+S_m.
\label{action}
\eeq
Here  $\k^2\equiv 8\pi G_{bare}$ is the bare gravitational constant
and $S_m$ denotes the action of matter (radiation and 
nonrelativistic particle). In our conventions, $\xi=1/6$ corresponds to 
the conformal coupling.

The basic idea is to utilize the dynamics during inflation.  
Several quintessence models require the fine-tuning of the initial 
conditions to account for the late-time domination of dark energy. 
The nonminimal coupling between the field and the Ricci scalar provides 
a natural mechanism to tune the field itself to a value which is required 
to achieve the cosmic acceleration today.


\paragraph*{Quintessence Axion.}

As an example, we consider a model based on  pseudo Nambu-Goldstone bosons
with the potential of the form \cite{pngb}: 
\beq
V(\phi)=M^4 (\cos(\k \phi)+1),
\eeq
where $M$ is a mass parameter. The smallness of $M$ may be related to 
 electroweak instanton effects or to a neutrino mass \cite{pngb}.
An attractive feature about this model 
is that the smallness of $M$ is technically natural and is protected from 
various corrections by symmetry: When the small mass is set 
to zero, it cannot be generated in any order of perturbation theory. 
On the other hand, the weakness is that the dynamics is 
significantly dependent on the initial condition: there does not exists 
a tracker solution \cite{trac} in this model, and 
$\phi$ must be tuned near 
the top of the potential $\phi \sim 0$ in order to account for the late-time 
domination. 

A nonminimal coupling can alleviate the situation. To demonstrate our point, 
we take the following functional form for $F(\phi)$ for simplicity:
\beq
F(\phi)={1\over 2}\xi\phi^2,
\eeq
where $\xi$ is a dimensionless parameter and we assume $\xi>0$ henceforth. 
The equations of motion in a FRW universe model are
\beqa
&&\ddot\phi +3H\dot\phi +V'+\xi \phi\left(\dot H +2H^2\right)=0,\\
&&3H^2=\k^2\left(\rho_B+{1\over 2}\dot\phi^2+V+
3\xi H\phi\left(H\phi+2\dot\phi\right)\right) =:\k^2(\rho_B+\rho_{\phi}),
\eeqa
where $V'=dV/d\phi$ and 
$\rho_B$ denotes the background (radiation/matter) energy density.

During inflation which is caused by other fields, the field $\phi$ 
acquires an effective (time-dependent) mass due to the nonminimal 
coupling $m_{eff}^2 \simeq\xi {\cal R}$ and is dynamically tuned toward 
$\phi=0$ (if $\xi>0$)\footnote{We thank Kazuya Koyama for pointing this out.} 
 which would subsequently correspond to the maximum of the potential 
induced by electroweak instanton effects or by a neutrino mass \cite{pngb}.
\footnote{The use of the nonminimal coupling for the restoration of symmetry 
in the inflationary stage was discussed in \cite{yokoyama}.} 
The evolution of $\phi$ during inflation is well approximated as 
$\phi(N)=\phi(N=0)\exp(-4\xi N)$ for $\xi \ll 1$, where $N$ is the e-folding 
number after the initial time.

In Fig. 1, we show an example of the numerical results of the time evolution 
of the energy densities. The scale factor, $a$, is normalized to unity today. 
In this example, we take $\k\phi=0.1$ and $\dot\phi=0$ at $z=10^{12}$ and 
$\xi=0.1$ and arrange $M$ to fix 
the density parameter of matter $\Omega_{M}=\k^2\rho_{M}/3H^2=0.4$ today. 
Interestingly, $\rho_{\phi}$ tracks  the radiation energy density 
$\rho_{R}$ during the radiation dominated epoch (RD)  
unlike minimal quintessence axion (see below for the details). 
In future $\phi$ will oscillate around a local minimum of the 
effective potential $V_{eff}(\phi)=V(\phi)+F(\phi){\cal R}$ similar to
 the minimal quintessence axion.  After the oscillations 
are damped, $\Omega_{\phi}$ decreases suddenly and settles down to a 
small but nonvanishing value, which is estimated $\simeq \xi \pi^2$ 
for $\xi \ll 1$. 
We note that the present Brans-Dicke 
parameter, $\omega_{BD}=(1-2\k^2F)/4\k^2F'^2$, is 
$\simeq 16000$ which is much larger than the current limit 
$\omega_{BD} >3500$ \cite{will}. Since $\k\phi$ remains small, $\k\phi \ll 1$, 
large $\omega_{BD}$ is attainable even with not so small $\xi$. The 
situation is different from tracker models where generically 
$\k\phi \simeq 1 $ today and thus $\xi$ is constrained to be very small, 
$\xi \siml 10^{-3}$ \cite{chiba}. 
We also note that extended quintessence 
models with positive $\xi$ can lead to a decrease in the primordial 
$^4{\rm He}$ abundance because of the decrease of the gravitational 
constant in the past \cite{bbn}.\footnote{We thank Kazunori Kohri for 
useful correspondence on this point.} However, the effect is found to be 
very small since $\xi\k\phi^2$ is small.

\begin{figure}[htdp]
  \begin{center}
  \leavevmode\psfig{figure=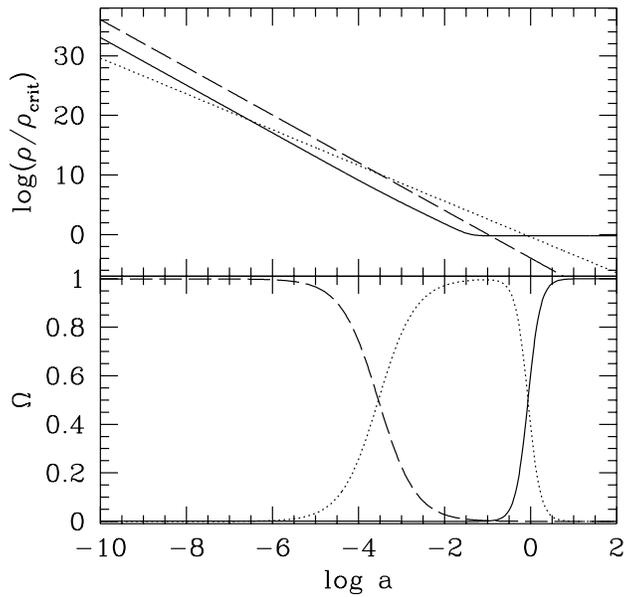,width=9cm}
  \end{center}
  \caption{The evolution of energy density for extended quintessence axion. 
The solid line is the energy 
density of dark energy, the dashed line is that of radiation, and the dotted 
line is that of matter. The lower panel shows the evolution of density 
parameters.}
  \label{fig:fig1}
\end{figure}

\paragraph*{Tracking without Tracking.}

Here we point out a novel tracking behavior during radiation dominated 
epoch (RD) in extended quintessence which is independent of the details of 
the potential. 
Consider the situation where 
the energy density of the scalar field defined usually for the minimal 
coupling case is much smaller than the energy density of radiation:
\beq
{1\over 2}\dot \phi^2+V \ll \rho_{R},
\eeq
and we assume that the curvature of the potential, $\sqrt{V''}$, is smaller 
than the Hubble parameter, $H$, during RD, which is the usually the case 
for most of quintessence models. 
Then the motion of the scalar field is almost frozen 
because the equation of motion of it during RD is 
the same as the usual minimal one.  
For $\xi \neq 0$, the energy density of the scalar field properly 
derived from the scalar field action 
$S_{\phi}=\int\sqrt{-g}[-(\nabla\phi)^2/2-
V(\phi)-F(\phi){\cal R}]$ becomes \cite{uzan}\footnote{The corresponding 
expression for the pressure is 
$p_{\phi}=\dot\phi^2/2-V-2\ddot F-4H\dot F-2F(2\dot H+3H^2)$. 
The conservation of the energy momentum tensor of $\phi$, 
$\dot\rho_{\phi}+3H(\rho_{\phi}+p_{\phi})=0$, reduces to the Klein-Gordon 
equation, $\delta S_{\phi}/\delta\phi=0$.}
\beq
\rho_{\phi}={1\over 2}\dot \phi^2+V(\phi) +6H\left(\dot F+HF\right)
\simeq 6H^2F.
\eeq
Since $F(\phi)$ is constant, this implies that the ratio of $\rho_{\phi}$ to 
$\rho_{R}$ is {\it constant} during RD although 
$\phi$ itself is {\it not} evolving. We shall call this behavior as 
``tracking without tracking''. As the universe becomes matter dominated 
the source term appears in the equation of motion of the scalar field, 
and the scalar field begins to move. 

In fact, in the above example, such a behavior of $\phi$ is found. 
In Fig. 2, we plot the time evolution of $\phi$. 
$\phi$ remains almost constant until the radiation-matter equality. 
The initial decrease  of $\phi$ after the radiation-matter 
equality is due to the curvature term induced by the nonminimal coupling. 
This is another dynamical tuning toward $\phi=0$ during the matter 
dominated epoch in addition to during inflationary epoch.

\begin{figure}[htdp]
  \begin{center}
  \leavevmode\psfig{figure=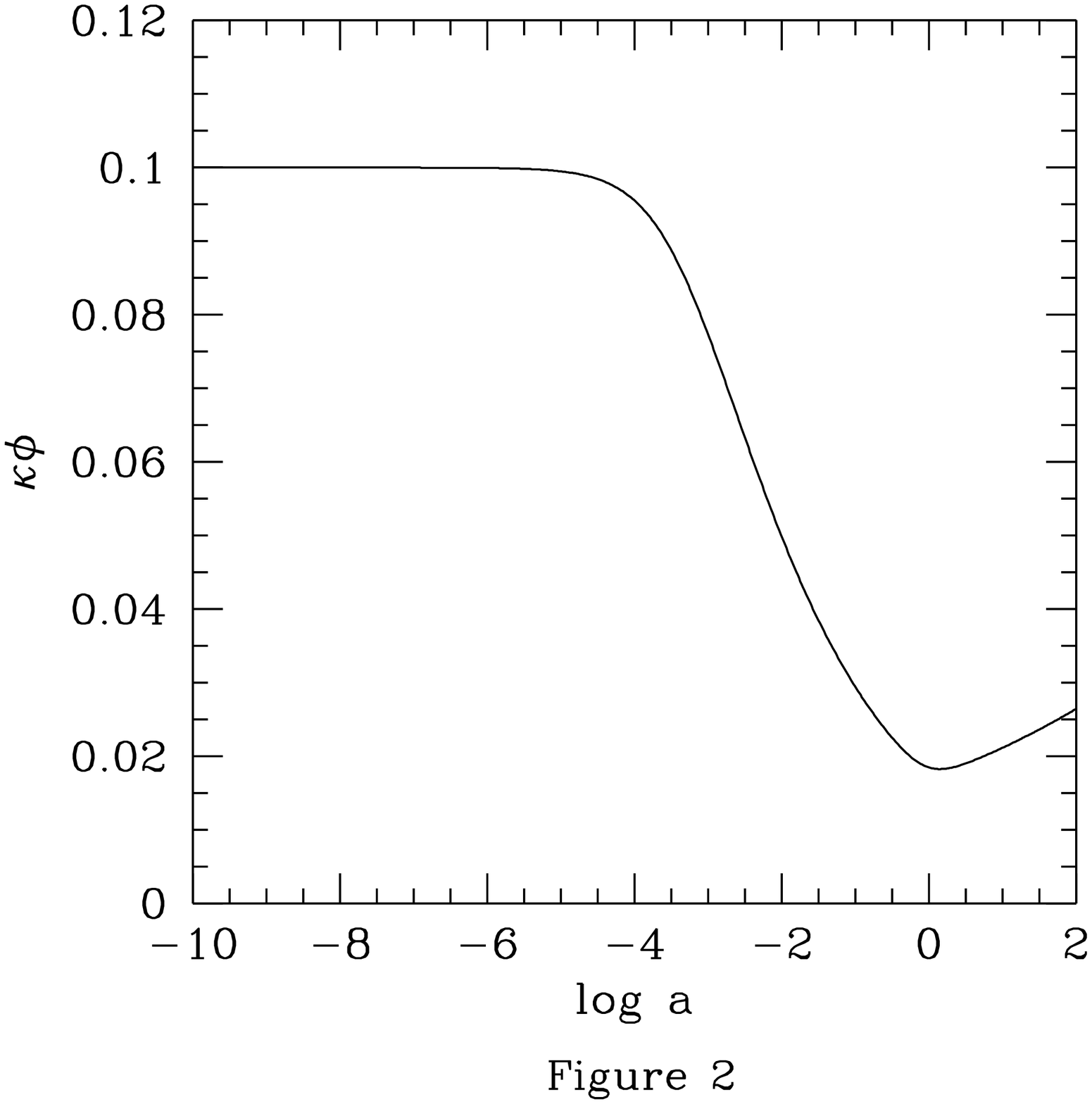,width=9cm}
  \end{center}
  \caption{The evolution of the scalar field for extended quintessence axion}
  \label{fig:fig2}
\end{figure}

We should emphasize that this behavior is independent of the details of 
the shape of a potential. The only constraint is that  
the energy density usually defined for $\xi=0$ is much smaller than 
the radiation energy density and that the curvature of the potential 
is  smaller than the Hubble parameter during RD. However, it is also 
to be noted 
that this ``tracking'' is not a dynamical attractor (or tracker \cite{trac}). 
In the extended quintessence axion, for example, the dynamics during RD 
is dependent on the initial conditions like the minimal quintessence axion 
simply because the equation of motion of the scalar field during RD is the 
same as that of minimal quintessence. 
``Tracking without tracking'' is rather kinematical tracker 
inherent in a wide class of extended quintessence.

\paragraph*{Exponential Potential.}

We also present another model with an exponential potential, 
$V\propto \exp\left(-\lambda \k\phi\right)$. 
Minimal quintessence models involving exponential potentials have been 
extensively studied \cite{exponential}. It has been shown that there 
exists the  attractor solution which depends only on $\lambda$, but the 
solution  tracks the matter/radiation energy density exactly 
in the same way and hence cannot account for the late-time domination of 
dark energy. 

The situation can be differrent in extended quintessence. A slight 
modification of $F(\phi)$ enables us to construct a model based on
an exponential potential which can account for the late-time domination of 
dark energy without the fine-tuning of parameters. For example, consider 
\beqa
&&F(\phi)={1\over 2}\xi(\phi-v)^2,\\
&&V(\phi)=\k^{-4}\exp\left(-\lambda \k\phi\right).
\eeqa
The role of the shift parameter $v$ is to reduce the energy scale of 
$V(\phi)$ by shifting $\phi$ without introducing small parameters in 
$V(\phi)$.\footnote{Introducing $v$ may nothing but paraphrase the smallness 
of the present energy scale of $V(\phi)$. 
But an interesting point may be that $v$ is at most only ${\cal O}(100)$ 
in $\k=1$ unit (bare Planck units) \cite{exp}. }
In Fig. 3, we show the evolution of the energy density of each component. 
In this example, we take $\xi=0.001,v=144\k^{-1},\lambda=2$. 
We note that the present Brans-Dicke parameter is 
$\omega_{BD} \simeq 7030$.

\begin{figure}[htdp]
  \begin{center}
  \leavevmode\psfig{figure=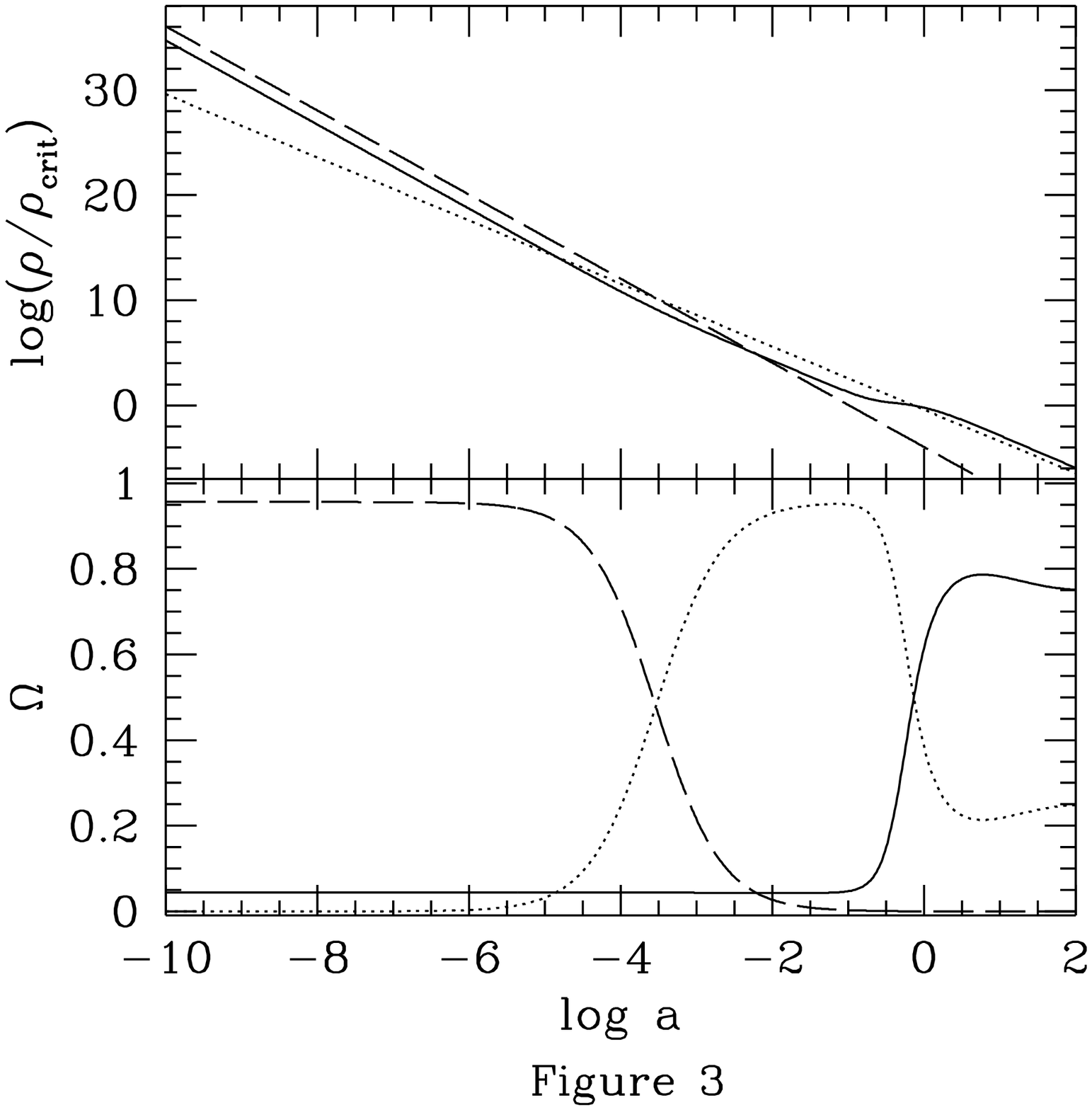,width=9cm}
  \end{center}
  \caption{The evolution of energy density for extended quintessence with 
an exponential potential. }
  \label{fig:fig3}
\end{figure}

\section{Summary}
In this paper, we have presented a scenario based on extended quintessence 
models to explain the late-time domination of dark energy without severe 
fine-tuning of initial conditions and model parameters. 
If the scalar field is nonminimally coupled to the curvature, inflationary 
dynamics  can provide the natural initial conditions required for  
the late-time domination of dark energy. 
Extended quintessence can subsequently track the radiation energy during 
the radiation dominated epoch even if minimal quintessence does not. 
This is due to  a novel tracking behavior (tracking without 
tracking) which is inherent in a wide class of extended quintessence models.
Further dynamical tuning occurs during matter dominated epoch. 
Two models have been presented; one is based on the pseudo Nambu-Goldstone 
bosons which is attractive because the small mass is technically natural, 
the other is based on an exponential potential which only uses the 
parameters in the potential roughly of ${\cal O}(1)$ in the Planck units. 
It will not be difficult to give other examples.

To conclude, extending quintessence to include nonminimal coupling to gravity 
 greatly extends the range of viable models for quintessence. 

\acknowledgments
This work was supported in part by a Grant-in-Aid for Scientific 
Research (No.13740154) from the Japan Society for the Promotion of Science.


\end{document}